\documentstyle[12pt,aasms4]{article}

\received{22 May 1998}
\accepted{1 June 1998}

\slugcomment{To appear in AJ, September 1998 issue}

\lefthead{M. S. Sahu}
\righthead{Absorption Lines in the NGC~1705 spectrum}

\newcommand{\kms}{$\mbox{km\,s}^{-1}$}

\begin{document}

\title{Interstellar Absorption Lines in the spectrum of the starburst
galaxy NGC~1705\footnotemark}

\author{M. S. Sahu}
\affil{NASA/Goddard Space Flight Center, Code 681, Greenbelt, MD~20771\\
and \\
National Optical Astronomy Observatories, 950 N. Cherry Avenue,
Tucson, AZ 8519-4933\\
msahu@panke.gsfc.nasa.gov}

\footnotetext{Based on observations taken with the NASA/ESA $\it{Hubble~
Space~Telescope}$, obtained at the Space Telescope Science Institute, which
is operated by the Association of the Universities for Research in Astronomy
under contract NAS 5-26555}


\begin{abstract}

\end{abstract}

I present a $GHRS$  archival study of the interstellar absorption lines in the line-of-sight to the H~{\sc i}-rich starburst dwarf galaxy NGC~1705, in the $\lambda$1170 to 1740 \AA\ range at $\sim$ 120 \kms\  resolution. 
The absorption features arising due to photospheric lines are 
distinctly different from the interstellar lines : the photospheric lines are weak, broad (equivalent widths $>$ 1\AA ), asymmetric and centred around the systemic LSR velocity of NGC~1705 ($\sim$ 610 \kms ).
The interstellar lines consist of three relatively narrow components at LSR velocities of --20, 260 and 540 \kms\ and
include absorption by neutral atoms (N~{\sc i} $\lambda$1200 \AA\ triplet and 
O~{\sc i} $\lambda$1302 \AA ) singly ionized atoms 
(Si~{\sc ii} $\lambda$1190, 1193, 1260, 1304 and 1526 \AA , S~{\sc ii} 
$\lambda$1253 \AA , C~{\sc ii} $\lambda$1334 \AA , 
C~{\sc ii}$^\ast$ $\lambda$1336 \AA , Fe~{\sc ii} $\lambda$1608 \AA\  and Al~{\sc ii} $\lambda$1670 \AA ) and atoms in higher ionization states (Si~{\sc iii} $\lambda$1206 \AA ,  Si~{\sc iv} $\lambda$1393, 1402 \AA\  and C~{\sc iv} $\lambda$1548, 1550 \AA ). The Si~{\sc iv} and C~{\sc iv} absorption features have 
both interstellar and photospheric contributions. \\

In an earlier study, Sahu \& Blades (1997) identified the absorption system at --20 \kms\  with  Milky Way disk/halo gas and
the 260 \kms\ system with a small, isolated high-velocity cloud HVC~487, which is probably associated with Magellanic Stream 
gas. The 540 \kms\ absorption system is associated with a kpc-scale expanding, ionized supershell centred on the superstar cluster NGC~1705-1.
The analysis presented in this paper consists of (1) a list of all interstellar 
absorption features with $>$ 3 $\sigma$ significance and their measured equivalent widths and (2) plots of the lines in the various
atomic species together with the results of non-linear least square
fit profiles to the observed data and (3) unpublished 21-cm maps from the 
Wakker \& van Woerden survey showing the large-scale H~{\sc i} distribution in the region
near the NGC~1705 sightline and HVC~487. Further, I report weak N~{\sc i} $\lambda$1200 \AA\  triplet absorption for the supershell component, which in the absence of dust depletion and ionization corrections implies a low N abundance. A low N
abundance for the supershell is consistent with an interpretation of nucleosynthetic enrichment by time-delayed ${\it primary}$ nitrogen production, the age estimate of 10 -- 20 $\times$ 10$^6$ years for the central superstar cluster NGC~1705-1 (Heckman and Leitherer, 1997) and the underabundance of Fe reported by Sahu and Blades. However, using the N~{\sc i} 1200\AA\  triplet alone to estimate the total N abundance could result in a severe underestimation of this quantity: although N does not deplete onto interstellar dust grains, photoionization and collisional ionization effects could increase the fraction of N found in higher ionization stages. Uncertainties in the total N abundance caused by photoionization and collisional ionization effects can only be addressed by future observations of the higher ionization lines, namely, N~{\sc ii} $\lambda$ 1084 \AA\  and N~{\sc iii} $\lambda$ 989 \AA .




%

\section{Introduction}

Absorption features in the line-of-sight to nearby galaxies provide a tool 
to study the properties of gas associated with three different environments, 
namely, gas in our Galaxy, gas associated with the background target galaxy and gas in the 
sightline associated with the intervening structures such as high velocity 
clouds (HVCs) in the Galactic halo and/or gas associated with the 
Local Group galaxies. The metal abundance of gas associated with the background galaxy can be determined from absorption spectra. This becomes particularly interesting when the background galaxy is a 
starburst galaxy: starburst galaxies release vast amounts of mass, energy, momentum and chemically-enriched gas into the 
surrounding intergalactic medium through supernova-driven galactic winds 
(e.g. Heckman et al., 1995). Deriving the metallicities of gas associated 
with outflows such as supershells, provides estimates of the chemical enrichment and star-formation history of the galaxy and an estimate of the extent starburst
galaxies (especially dwarf galaxies) pollute the local intergalactic 
medium.\\

NGC~1705, a nearby (distance $\sim$ 6~Mpc) H~{\sc i}-rich dwarf galaxy,
contains the principal constituents of a starburst including a 
kpc-scale expanding shell (Meurer et al., 1992) and a luminous 
central superstar cluster, NGC~1705-1 (Sandage, 1978 \& Melnick et al., 1995). 
The UV spectrum of NGC~1705-1 obtained using the Goddard High Resolution Spectrograph ($GHRS$) on the Hubble Space Telescope shows strong absorption lines which are interstellar in origin (Heckman \& Leitherer, 1997).
Using the same $GHRS$ spectra, Sahu and Blades (1997; hereafter SB~97) reported the presence of three principal absorption systems at LSR velocities --20 \kms\ , 260 \kms\ and 540 \kms\ . They argued that the absorption features are interstellar rather than stellar in origin 
based on a 
comparison of the Si~{\sc ii}
$\lambda$1526.71 \AA\  with the photospheric C~{\sc iii} 
$\lambda$1175.7 \AA\  absorption feature.
The --20 \kms\ absorption line system was identified with  Milky Way disk/halo gas 
and the 260 \kms\ system with an isolated high-velocity cloud HVC~487 
associated with Magellanic Stream  gas. The 540 \kms\ absorption 
system was identified with a blue-shifted emission component associated 
with a kpc-scale expanding supershell of ionized gas centered on NGC~1705-1.
The most striking feature of the 540 \kms\ absorption line system is strong Si~{\sc ii} and Al~{\sc ii} absorption but weak  Fe~{\sc ii} $\lambda$1608 \AA\  absorption. SB~97 interpreted the
low intrinsic Fe abundance of the supershell component in the context of
supernova-driven galactic wind evolution of dwarf galaxies. \\

In this paper, I extend the study of SB~97 to other atomic species and
provide a list of interstellar absorption lines towards NGC~1705-1 in the 
$\lambda$ 1170 to 1740 \AA\ range and their measured equivalent widths. Plots of the absorption
in various atomic species together with the results of non-linear least square
fit profiles are presented. Further arguments confirming the 
interstellar nature of the absorption features are outlined. The 540 \kms\ component associated with the supershell in NGC~1705 shows weak N~{\sc i} $\lambda$ 1200\AA\ triplet absorption which is further discussed in $\S$ 7. \\

\section{The Observations and data reduction}

The present study is based on two $GHRS$ spectra of NGC~1705-1 taken 
through the 
small science aperture with the G140L grating, which were retrieved from the HST Data 
Archives. The first spectrum extends from $\lambda$1170 to 1462 \AA\  and the second one from $\lambda$1454 to 
1740 \AA . The velocity resolution of spectra are $\sim$ 140 \kms\ at $\lambda$ 1200 \AA\     
and $\sim$ 100 \kms\ at $\lambda$1700 \AA\   and the signal-to-noise ratio (S/N) in the 
continuum ranges from $\sim$ 7:1 at 1670 \AA\    to 16:1 at 1400 \AA . These archival 
$GHRS$ spectra represent the highest S/N and highest velocity 
resolution UV spectra ever taken towards NGC~1705-1. The reader is referred to SB~97 for details 
of the $GHRS$ data reduction procedures employed. \\

\section{Presentation of the data}

Between $\sim$ 1300 and 1700 \AA , the continuum in the $GHRS$ spectra is 
smooth and slowly-varying  (Figure 1 of Heckman \& Leitherer 1997, shows the 
flux-calibrated $GHRS$ spectrum of NGC~1705-1). The spectra were normalized by 
fitting low-order polynomial functions and the continuum-normalized spectra 
between $\lambda$1175 to 1740\AA , are presented in Figures 1 (a -- f). 
All absorption features with $>$ 3$\sigma$ significance have been identified
and the identifications are indicated in the figure with the following notation: $\times$ indicates a photospheric absorption feature, $\otimes$
indicates a blend and an unidentified line is indicated as `UL'. All features
that have an interstellar origin (refer $\S$ 4) are entered in Table 2.
Table 2 contains the equivalent measurements of the interstellar absorption 
lines at the --20 \kms, 260 \kms\  and  540 \kms\  (LSR velocities have been used
throughout this paper). Equivalent width measurements of blended lines
are uncertain and have not been listed in this table. The laboratory wavelengths
and oscillator strengths are listed and the references are indicated in
the footnote.
The last part of the data presentation consists of plots of the interstellar
absorption lines in the various atomic species together with the results of the non-linear least square profile fits (presented in $\S$ 5, 6 and 7).\\

\notetoeditor{Figure 1 (a -- f)}\\

\section{Further evidence for interstellar origin of the absorption lines}

Absorption features in the spectra of NGC~1705-1 can originate in (1) the 
interstellar medium of our Galaxy (2) in the intervening medium (3)
in the interstellar medium of NGC~1705, or (4) in the stellar atmospheres
of stars contributing to the UV flux of the galaxy. 
Previous studies of the interstellar absorption lines towards NGC~1705
have been made by York et al. (1990), using $IUE$ spectra. They used both low-dispersion (velocity resolution $\sim$ 1000 \kms) as well 
as higher resolution $IUE$ spectra (resolution $\sim$ 30 \kms) towards 
NGC~1705, 
to argue that the absorption lines seen in these spectra were primarily interstellar (also refer Heckman \& Leitherer, 1997).
SB~97 used the higher S/N $GHRS$ data and 
confirmed the conclusions of York et al. (1990) that the UV absorption
components seen in the NGC~1705 spectra are interstellar in origin.\\

The interstellar origin of the absorption lines is further supported in this paper by a compilation of all the photospheric lines present in the $GHRS$ spectra
(namely, C~{\sc iii} $\lambda$1175 \AA , C~{\sc iii} $\lambda$1247 \AA , 
Si~{\sc iii} $\lambda$1294, 1296 \AA\ 
and Si~{\sc iii} $\lambda$1417 \AA ). The photospheric profiles are shown in Figure 2
and the LSR recession velocity of NGC~1705 ($\sim$ 610 \kms ) is indicated by 
an arrow. In all the cases, the photospheric absorption lines are weak, broad 
with FWHM $\sim$ 200 to
300 \kms\ and have asymmetric blue-ward wings characteristic of stellar 
mass loss. \\

\notetoeditor{Figure 2}\\

The Si~{\sc ii} $\lambda$1526.71 \AA\  line has three narrow components :  
a strong feature centred at --20 \kms \ , a relatively weak 
feature centred at  260 \kms\ , and a strong feature centred at  540 \kms\ . 
All the other resonance lines in the spectra show similar 
velocity structure. Their
velocities are not consistent with the systemic velocity of NGC~1705, 
nor do any of the profiles show asymmetric or P-Cygni shapes characteristic 
of stellar mass loss. Hence, apart from the broad, asymmetric 
C~{\sc iii} 1175 \AA , C~{\sc iii} 1247 \AA , 
Si~{\sc iii} 1294, 1296 \AA\ 
and Si~{\sc iii} 1417 \AA\  lines, the other lines can be attributed to an interstellar origin without ambiguity. 
The C~{\sc iv} and Si~{\sc iv} lines show features common to photospheric and 
interstellar lines: the three narrow components superposed on
a broad asymmetric component (refer $\S$5.3).

\section{Interstellar absorption features}

A simple curve-of-growth (cog) analysis for the three components 
at --20 \kms ,  260 \kms\ and 540 \kms\ results in best-fitting 
$b$-values of $\sim$ 25 \kms , $\sim$ 50 \kms\ and $\sim$ 30 \kms\  
respectively. The optical depths at the center of the line, $\tau$$_\circ$, for the three absorption systems are listed in Table 1; only lines with  
$\tau$$_\circ$ $\leq$ 1 can be used for column density 
determinations ; stronger lines provide lower limits.\\

\subsection{The neutral species}

Narrow interstellar absorption lines of the N~{\sc i} $\lambda$ 1200 \AA\  triplet
  were detected in the spectra. The  
N~{\sc i} line is detected at --20, 260 and 540 \kms . The N~{\sc i} 
$\lambda$1200 \AA\  triplet consists of three lines at 1199.6, 1200.2 and 1200.7 \AA\  which have approximately the same strengths. The three lines 
of the N~{\sc i} triplet are blended on the $GHRS$ spectra. 
In spite of this blending, it is apparent from Figure 3, that the absorption features at --20 and 260 \kms\ are moderately strong but conspicuously weak at
540 \kms\ (refer $\S$ 7).
The O~{\sc i} $\lambda$1302.2 \AA\ lines are detected but are heavily blended with the Si~{\sc ii} $\lambda$ 1304.4\AA\  line.
Broad damped Lyman-$\alpha$ absorption is also detected (Figure 3).
The damped Lyman-$\alpha$ absorption line
gives a reasonably accurate measurement of the neutral hydrogen column densities, but
because the absorption is strong there is very little information on the velocity
components of the H~{\sc i} gas. The damped Lyman-$\alpha$ line was fitted 
with a theoretical Lyman-$\alpha$
damping profile. The fitting was intiated with two absorber 
velocities at --20 \kms\ and 540 \kms\ but the resulting profiles were unsatisfactory.
The best fit was obtained for absorber velocities of 104 and 678 \kms\ and column densities log$N$ = 20.1 and 19.6 cm$^{-2}$ respectively.
The H~{\sc i} gas at $\sim$ 678 \kms\ is probably H~{\sc i} gas associated with 
the dwarf galaxy. Information regarding the H~{\sc i} gas associated with the 
supershell at 540 \kms\ is impossible to disentangle. \\

\notetoeditor{Figure 3}\\

\subsection {Singly-ionised species}

Figure 4 is a plot of all the single ionized species detected in the 
$GHRS$ spectra. Two resonance lines of singly ionized carbon, C~{\sc ii} $\lambda$ 1334.53 \AA\  and C~{\sc ii}$^\ast$ $\lambda$ 1335.71\AA\ are detected. The velocity resolution
is not sufficient, resulting in blending of these two lines. In addition there is a contribution from a broad, photospheric C~{\sc ii} $\lambda$ 1334.53 \AA\ 
absorption feature, making it difficult to derive accurate equivalent widths.\\

Strong Al~{\sc ii} $\lambda$ 1670.7 \AA\ 
absorption is present at both --20 and 540 \kms\ and has been extensively 
discussed in SB~97. 
The best observed species in the $GHRS$ spectra is Si$^+$
with five definite absorption line detections at $\lambda$1190.24, 1193.29
1260.42, 1304.37 and 1526.71\AA . For the --20 and 540 \kms\ components, 
with the exception of the $\lambda$1304.37 \AA\ line, all the other 
Si$^+$ lines have $\tau$$_\circ$ $>$ 1 and only lower limits to the column densities can be derived for these components. The 260 \kms\ component corresponding to HVC~487
has an optical depth of $\tau$$_\circ$ $\leq$ 1 at the centre of the line and has been used by SB~97
to derive a column density of N(Si) = 4$\times$10$^{13}$ cm $^{-2}$ and
a metal abundance estimate (Si/H) $>$ 0.6(Si/H)$_\odot$ (see $\S$ 6)\\

S~{\sc ii} $\lambda$1259.52 \AA\ absorption is present at the 3$\sigma$ level
at --20 and 540 \kms\ . Only one Fe~{\sc ii} line at 1608.45\AA\  is covered
by the $GHRS$ spectra. Strong Fe~{\sc ii} absorption is detected at --20 \kms\
but only weak Fe~{\sc ii} absorption is detected at 540 \kms\ . The low Fe/Al
abundance ratio has been extensively discussed by SB~97. No Ni$^+$ $\lambda$
1317.22\AA\  absorption is detected in the spectra at the 3$\sigma$ level. 
\notetoeditor{Figure 4}\\

\subsection {Higher ionization species}

The Si~{\sc iii} $\lambda$1206.50 \AA\  and the doublet resonance absorption lines
of Si~{\sc iv} and C~{\sc iv} are detected and shown in Figure 5.
The C~{\sc iv} and Si~{\sc iv} lines show features common photospheric and 
interstellar lines: the three narrow components superposed on
a broad asymmetric component. The Si~{\sc iv} lines, in particular, are
heavily contaminated by broad, asymmetric photospheric component. 
No attempt has 
been made to subtract this photospheric component. The 
C~{\sc iv} and Si~{\sc iv} observed profiles are shown 
in Figure 5, although no equivalent measurements have been made due to 
the large uncertainties caused by photospheric contamination.\\

\notetoeditor{Figure 5}\\

\section{The high velocity cloud HVC~487}

The NGC~1705-1 sight-line intercepts a region which 
is approximately 10$^\circ$ away from the outermost H~{\sc i}
contours that envelops both the Large and Small Magellanic Clouds 
on 21-cm maps (Mathewson \& Ford, 1984). The 260 \kms\ absorption component 
seen in the UV spectra is associated with the isolated high velocity cloud,
HVC~487 on the 21-cm map, based on velocity agreement (SB~97). 
The location of HVC~487 with respect to the NGC~1705 
sightline, the Magellanic Clouds and the Magellanic Stream is shown in 
Fig 6. The data presented in this figure is 
an unpublished H~{\sc i} map of this region obtained from 
the all-sky HVC survey of Wakker and van Woerden (1991).
The (Si/H) obtained by SB~97 for HVC~487 is 0.9 (Si/H)$_\odot$,
close to the metal abundance estimates for the Magellanic Stream
(Lu et al. 1994). The velocity agreement together with the agreement
in metal abundances, point to a Magellanic Stream origin for HVC~478.

\notetoeditor{Figure 6}\\

\section{Low N abundance for the NGC~1705 supershell component?}

Galactic halos (e.g. Bowen et al. 1996) and gas-rich dwarf galaxies (York et al. 1986) have been proposed as giving rise to QSO absorption line systems (QSOALS). Star-formation induced gas outflows like supershells, emanating from starburst galaxies are another viable cause of QSOALS because of the following reasons: 
(1) they produce line strengths, ionizations and velocity structures similar to gas-rich dwarf galaxies (2) supershells from starburst galaxies extend to several kpc away from the nucleus of the galaxy thereby increasing the galaxy
absorption cross section and (3) shell-like structures are an ubiquitous component of the ISM of galaxies, for example, the H~{\sc i} supershells in the 
Milky Way (Heiles, 1990) and M~101 (Kamphuis et al. 1991). The supershell component in the NGC~1705 sightline is a possible low redshift analogue to QSOALS, making it an excellent sightline to study the age,  
chemical enrichment and star-formation history of a dwarf galaxy, namely,
NGC~1705.\\

The N/Si and N/0 abundance ratios are good indicators of the age and
star-formation history of a galaxy. {\it Primary} nitrogen is produced in the AGB phase 
of the evolution of intermediate-mass stars (3 - 8 M$_\odot$) and the delayed 
release of {\it primary} N causes a galaxy to zig-zag on N/O versus O/H plots, as it evolves. This technique has been pioneered by Pagel and collaborators (e.g. Pagel et al., 1992): the position of a dwarf
galaxy on a N/O versus O/H plot determines its star-formation history
(refer Fig. 7, Garnett, 1990).
If star-formation proceeds in bursts separated by quiescent periods, 
the N/O increases with time for a fixed O/H. For example, a dwarf galaxy with 
mass 10$^8$M$_\odot$ turns 1\% of its mass into massive stars in a 
burst quickly increasing the oxygen
abundance of the whole galaxy by a factor of 6 and moving on a
45$^\circ$ line to the lower right. Ejection of nitrogen from 
intermediate-mass stars then increases N/O at constant O/H 
after the starburst has
faded (Edmunds \& Pagel, 1978, Renzini \& Voli, 1981).
A second burst of the same size again moves the galaxy to the lower
right, but by a smaller amount.\\

This technique has been extended to damped Lyman-$\alpha$ galaxies
by Pettini et al. (1995) who first reported that the $z$$_{abs}$ =2.27936
absorption system in the QSO~2348--147 sightline had significantly
underabundant N indicative of {\it primary} N production. The
O~{\sc i} interstellar lines are usually heavily saturated so the N/Si
abundance ratio was used as a
substitute for the N/O abundance ratio. Lu et al. (1998) extended 
this study to 15 damped Ly-$\alpha$ galaxies and interpreted
the scatter in the N/Si versus Si/H plots as support for the
delayed release of primary N production in intermediate-mass stars.\\

This same technique can be applied to the supershell component to determine
the age and star-formation history of NGC~1705.
Figure 7 shows a plot of relative intensities
against LSR velocities in the N~{\sc i} 1200 \AA\  line (shown as a solid
histogram). Plotted over is the Si~{\sc ii} 1526\AA\  observed profile 
(as dashed histogram). The N~{\sc i} feature at 540 \kms\ is much
weaker than the corresponding Si~{\sc ii} feature. 
In the diffuse ISM in our Galaxy, little or no N is accreted onto interstellar
grains (Jenkins, 1987, Sofia et al. 1994), 
hence depletion onto dust cannot account for the weak N~{\sc i} absorption
in the supershell component. \\ 

The age estimates for the superstar cluster NGC~1705-1 
of 10 to 20 Myr, estimated by Heckman \& Leitherer (1997), provide a
natural explanation for the observed column densities and the underabundance
of Fe (Type Ia SN product) and N (produced by intermediate stars) compared
to Si and Al (produced by Type II SNe) implying the shell could have been
formed by the first generation of stars ($<$ 10$^8$ years) and is therefore
not contaminated by products from either intermediate stars or Type Ia SNe
(Wheeler et al. 1989).
Although a nucleosynthetic enrichment interpretation for the low
N abundance due to {\it primary} N production is consistent with the
age of NGC~1705-1 and the various timescales, there are two important
uncertainties: (1) Using the photoionization code $CLOUDY$ (Ferland, 1993),
a spectral type B3V for the ionizing source NGC~1705-1 (Melnick et al. 1985)
and a plane-parallel cloud, the N~{\sc i}/N~{\sc ii} is estimated to be 
$\sim$ 2,
indicating a significant amount of nitrogen could be in higher ionization
stages,  and (2) the weak N~{\sc i} for the shell component could be the result 
of collisional ionization effects as suggested by Trapero et al. (1996).\\

The uncertainties in the total N abundance caused by photoionization and collisional ionization effects cannot be addressed by the $GHRS$ spectra
discussed in this paper. Observations of the higher 
ionization stages of nitrogen like, N~{\sc ii} $\lambda$ 1084\AA\  and 
N~{\sc iii} $\lambda$ 988\AA\   (observable with $FUSE$) will allow an assessment
of the effects of photoionization and collisional ionization. The main advantage of investigating N/Si abundance ratio in the supershell component over the damped Lyman-$\alpha$ galaxies is that there is no need to make assumptions regarding the ionization source or the
absorber. The properties and age of the ionizing central star
cluster NGC~1705-1 and the supershell are well-studied (e.g. Meurer et al.
1992) and the significance of collisional ionization effects can be assessed.\\

\notetoeditor{Figure 7}\\

\section {Summary}

An analysis of $GHRS$ spectra in the  $\lambda$1170 to 1462 \AA\  range towards NGC~1705-1, the superstar cluster in the nearby dwarf starburst galaxy,
NGC~1705 at $\sim$  120 \kms\ is presented. These  
$GHRS$ spectra represent the highest S/N and highest velocity 
resolution UV spectra ever taken towards NGC~1705-1. The main results of this analysis are:\\
1) The photospheric absorption lines C~{\sc iii} $\lambda$1175 \AA , C~{\sc iii} $\lambda$1247 \AA , Si~{\sc iii} $\lambda$1294, 1296 \AA\ 
and Si~{\sc iii} $\lambda$1417 \AA\ are weak, broad and asymmetric and 
distinctly different from the interstellar absorption lines. Interstellar
absorption is detected in (i) neutral atoms (N~{\sc i} $\lambda$1200 \AA\ triplet and O~{\sc i} $\lambda$1302 \AA ) singly ionized atoms 
(Si~{\sc ii} $\lambda$1190, 1193, 1260, 1304 and 1526 \AA , S~{\sc ii} 
$\lambda$1253 \AA , C~{\sc ii} $\lambda$1334 \AA , 
C~{\sc ii}$^\ast$ $\lambda$1336 \AA , Fe~{\sc ii} $\lambda$1608 \AA\  and Al~{\sc ii} $\lambda$1670 \AA ) and atoms in higher ionization states (Si~{\sc iii} $\lambda$1206 \AA ,  Si~{\sc iv} $\lambda$1393, 1402 \AA  and C~{\sc iv} $\lambda$1548, 1550 \AA . The Si~{\sc iv} and C~{\sc iv} absorption features have both interstellar and photospheric contributions.\\
2) All absorption features with $>$ 3$\sigma$ significance have been identified and equivalent measurements of the unblended lines are listed in Table 1.
Plots of the interstellar absorption lines in the various atomic species together with the results of the non-linear least square profile fits
are presented in Figs. 3, 4 and 5. Three interstellar absorption systems at LSR velocities of --20 \kms, 260 \kms\  and  540 \kms\  are detected : the --20 \kms\ absorption line system is associated  with  Milky Way disk/halo gas and the 260 \kms\ system with an isolated high-velocity cloud HVC~487 associated with Magellanic Stream  gas. The 540 \kms\ absorption 
system is associated with a blue-shifted emission component of
a kpc-scale expanding supershell of ionized gas centered on NGC~1705-1.\\
3) Large-scale H~{\sc i} 21-cm maps of the regions near the NGC~1705-1
sightline and Magellanic Stream region show the isolated high velocity cloud,
HVC~487 (Fig. 6). Based on velocity and metallicity agreement, HVC~487 is probably
associated with Magellanic Stream gas.\\ 
4) The 540 \kms\ absorption line system has strong Si~{\sc ii} and Al~{\sc ii} absorption but weak  Fe~{\sc ii} $\lambda$1608 \AA\  absorption which
Sahu \& Blades (1997) interpreted in the context of
supernova-driven galactic wind evolution of dwarf galaxies. In this paper,
the apparent low N abundance of this supershell component is reported.
Although a nucleosynthetic enrichment interpretation for the low
N abundance due to {\it primary} N production is consistent with the
age of NGC~1705-1 of 10 to 20 Myr, deriving the N abundance using the
N~{\sc i} $\lambda$1200 \AA\ triplet has uncertainties due to
photoionization and collisional ionization effects. These uncertainties can only be addressed by future observations of the higher ionization lines N~{\sc ii} $\lambda$ 1084 \AA  and N~{\sc iii} $\lambda$ 989 \AA .

\acknowledgements

I wish to thank Bart Wakker for providing the unpublished 21-cm data, 
Jim Lauroesch for discussions and an anonymous referee for 
helpful comments.

\clearpage 
\begin{deluxetable}{cclcccccccc}
\small
\footnotesize
\scriptsize
\tablecaption{Line list and equivalent width measurements of the interstellar absorption lines towards NGC ~1705 in the $\lambda$ 1170 to 1740 \AA\ range\label{tbl-1}}
\tablewidth{0pt}
\tablehead{
\colhead{Species} & \colhead{$\lambda$$_{vac}$} & \colhead{$f$}  & \colhead{} &
\colhead{EW (m\AA  )} & \colhead{} &
  & \colhead{$\tau$\tablenotemark{a}$_{\circ}$} & \colhead{} & \colhead{Comments} & \colhead{Ref\tablenotemark{d}} \\
\colhead{} & \colhead{(\AA )} & \colhead{} & 
\colhead{--20 \kms\ } & 
\colhead{260 \kms\ } & \colhead{540 \kms\ }  & \colhead{--20} & \colhead{260} & \colhead{540} &\colhead{} &\colhead{}
} 
\startdata
Si~{\sc ii} & 1190.42  & 0.2502\tablenotemark{b} &  $449^{+26}_{-22}$ & $88^{+10}_{-12}$ & $\it{blend}$ &3.9 &0.25 & \nodata & blend Si~{\sc ii} 1193 &(1)\nl
Si~{\sc ii} & 1193.29 & 0.4991\tablenotemark{b} & $\it{blend}$ & $101^{+21}_{-23}$ & $636^{+35}_{-94}$ & \nodata & 0.2 & $>$ 20 &\nodata &(2)\nl
N~{\sc i}   & 1199.55  & 0.1328\tablenotemark{b} &  $330^{+40}_{-36}$ & $474^{+93}_{-27}$ & $166^{+34}_{-25}$  & 1.3 & 1.1 & 0.5 &\nodata &(2)\nl
Si~{\sc iii} & 1206.50 & 1.669\tablenotemark{b}  & $698^{+37}_{-32}$ & $532^{+45}_{-52}$ & 1015$^{+112}_{-70}$ &$>$20 & 0.8 & $>$ 20 &\nodata & (2)\nl
S~{\sc ii} & 1253.81 & 0.01088\tablenotemark{b}& $151^{+66}_{-18}$ & \nodata & $181^{+86}_{-56}$ & 0.4 & \nodata & 0.26 &\nodata & (1)\nl 
Si~{\sc ii} & 1260.42  & 1.007\tablenotemark{b} &  $\it{blend}$     & $\it{blend}$    & $704^{+14}_{-22}$ & \nodata & \nodata & $>$20 &\nodata &(1)  \nl
O~{\sc i}  &  1302.17 & 0.0489\tablenotemark{b} & $\it{blend}$ & $\it{blend}$ & $\it{blend}$ & \nodata & \nodata & \nodata &\nodata & (2) \nl
Si~{\sc ii} & 1304.37  & 0.086\tablenotemark{b} & $\it{blend}$  & $157^{+47}_{-34}$ & $370^{+27}_{-19}$ & \nodata & 0.42 & 1.85 &\nodata &(1)\nl
C~{\sc ii} & 1334.53 & 0.01277\tablenotemark{b} & $\it{blend}$ & $\it{blend}$ & $\it{blend}$ & \nodata & \nodata & \nodata & \nodata &(2) \nl
C~{\sc ii}$^{\ast}$ & 1335.71 & 0.01149\tablenotemark{b} & $\it{blend}$ & $\it{blend}$ & $\it{blend}$ & \nodata & \nodata & \nodata & \nodata &(2) \nl
Si~{\sc iv} & 1393.76 & 0.5140\tablenotemark{b} & $\it{blend}$ & $\it{blend}$ & $\it{blend}$ & \nodata & \nodata & \nodata &blend with photospheric Si~{\sc iv} & (2) \nl
Si~{\sc iv} & 1402.77 & 0.2553\tablenotemark{b} & $\it{blend}$ & $\it{blend}$ & $\it{blend}$ & \nodata & \nodata & \nodata &blend with photospheric Si~{\sc iv} & (2) \nl
Si~{\sc ii} & 1526.71  & 0.110\tablenotemark{c} & $399^{+31}_{-30}$ & $239^{+62}_{-28}$ & $661^{+21}_{-19}$ & 5.5 & 0.4 & 8.6 & \nodata &(1)\nl
C~{\sc iv} & 1548.19 & 0.1908\tablenotemark{b} & $\it{blend}$ & $\it{blend}$ & $\it{blend}$ & \nodata & \nodata & \nodata & blend with photospheric C~{\sc iv} & (2) \nl
C~{\sc iv} & 1550.77 & 0.0.09522\tablenotemark{b} & $\it{blend}$ & $\it{blend}$ & $\it{blend}$ & \nodata & \nodata & \nodata  & blend with photospheric C~{\sc iv} & (2) \nl
Fe~{\sc ii} & 1608.45 & 0.062\tablenotemark{b}& $314^{+63}_{-20}$ & \nodata & $286^{+15}_{-44}$ & 2.3 & \nodata & 0.5 & \nodata & (1)\nl
Al~{\sc ii} & 1670.79 & 1.88\tablenotemark{b} & $508^{+70}_{-81}$ & \nodata & $624^{+45}_{-20}$ & 6.4& \nodata &  11.4 & \nodata &(1)\nl
\nl
\enddata


\tablenotetext{a} {optical depth at the center of the line for the --20, 260 and 540 \kms\ components} 
\tablenotetext{b} {Morton, 1991}
\tablenotetext{c} {Spitzer \& Fitzpatrick, 1993}
\tablenotetext{d} {References: (1) = SB~97~~~(2) = this work}
\end{deluxetable}

\clearpage

%
%

\figcaption[fig1.epsi]{(a--f). The continuum-normalized $GHRS$ spectra towards NGC~1705 from $\lambda$1173 to $\lambda$1740 \AA . All interstellar
absorption features with
$>$ 3$\sigma$ significance are labelled in the figure :
$\times$ indicates an absorption feature with photospheric origin,
$\otimes$ indicates a blend and unidentified lines are indicated by UL. All absorption features with an interstellar origin ($\S$ 4) are entered in
Table 1.\label{1}}

\figcaption[fig2.epsi]{The photospheric absorption features in the $GHRS$ spectra are plotted against LSR velocity. Note the broad,
asymmetric nature of the line profiles. The systemic velocity of $\sim$ 610
\kms\ derived from H~{\sc i} observations of NGC~1705,
is shown by arrows. Compare the broad, photospheric absorption features with 
the three narrow interstellar absorption components seen in Figure 4 
at --20 \kms\ , 260 \kms\ and 540 \kms\ , shown with arrows on 
the Si~{\sc ii} $\lambda$1526.71 \AA  profile. \label{2}}

\figcaption[fig3.epsi]{(a-c) Normalized interstellar absorption profiles in the neutral atomic species are plotted as histograms against LSR velocities.
Results of the non-linear least square fits are plotted as continuous lines.
The model fit for the N~{\sc i} profile includes contributions from the 
$\lambda$ 1199.55, 1200.22 and 1200.71\AA\ lines of the N~{\sc i} triplet. The O~{\sc i} absorption features are blended with the Si~{\sc ii} absorption features and no model-fitting was done due
to confusion. The damped Lyman-$\alpha$ absorption line is shown in panel
(c) together with the theoretical damped profile.\label{3}}

\figcaption[fig4.epsi]{Normalized interstellar absorption profiles in the singly ionized atomic species are plotted as histograms against LSR velocities.
Results of the non-linear least square fits are plotted as continuous lines. 
 Note the strong absorption at 540 \kms\ in the Si~{\sc ii}, 
S~{\sc ii} and Al~{\sc ii} lines. The Fe~{\sc ii} absorption at this 
velocity, on the hand, is relatively weak (Sahu \& Blades, 1997)  \label{4}}

\figcaption[fig5.epsi]{Normalized interstellar absorption profiles in the higher ionization stages are plotted as histograms against LSR velocities with the 
results of the non-linear least square fits plotted as continuous lines. 
The  Si~{\sc iv} and C~{\sc iv} lines show three narrow interstellar features superposed on a broad, photospheric component.
No profile fitting has been done for the Si~{\sc iv} and C~{\sc iv} lines
because of the uncertainties caused by photospheric contamination.
\label{5}}

\figcaption[fig6.epsi]{Plot showing H~{\sc i} 21-cm distribution in the region 
near the NGC~1705 sightline in galactic coordinates. The contour levels are represent 
N(H~{\sc i}) = 2 $\times$ 10$^{18}$,
5 $\times$ 10$^{18}$, 1 $\times$ 10$^{19}$, 5 $\times$ 10$^{19}$, 
1 $\times$ 10$^{20}$ and 5 $\times$ 10$^{20}$ cm$^{-2}$. The grey-scale 
represents the velocity field : white = 100 to 150 \kms\ , 
light grey = 150 to 200 \kms\ , medium grey = 200 to 250 \kms\ , 
dark grey = 250 to 300 \kms\ and black = 300 to 350 \kms\ .
 Si~{\sc iv} and C~{\sc iv} lines
Based on velocity agreement, the 260 \kms\ 
absorption component seen on the $GHRS$ spectra is associated with the 
small isolated high-velocity cloud HVC~487 (see $\S$6). The metallicity 
obtained 
for this component and therefore HVC~487, is similar to the Magellanic 
Cloud and Magellanic Stream indicating close association. \label{6}}

\figcaption[fig7.epsi]{Normalized interstellar absorption profiles of N~{\sc i}  are plotted as solid histograms against LSR velocities. The Si~{\sc ii}
$\lambda$ 1526.71\AA\ profile
is plotted over as a dotted histogram. Note the weak N~{\sc i} absorption 
at 540 \kms\ in comparison to the Si~{\sc ii} absorption (see $\S$ 7). 
\label{7}}


\begin{thebibliography}{}

\bibitem [Bowen 1996]{bow96} Bowen, D. V., Blades, C. J. Pettini, M. 1996,
\apj, 464, 141

\bibitem [Edmunds 1978]{ed96} Edmunds, M. G. \& Pagel, B. E. J. 1978,
\mnras, 185, 77

\bibitem [Ferland 1993]{fer93} Ferland, G. J. 1993, University of Kentucky Center for 
Computational Science Internal Report

\bibitem [Garnett1990]{g90} Garnett, D. R. 1990, \apj, 363, 142

\bibitem [Heckman 1995]{h95} Heckman, T.M., Dahlem, M., Lehnert, M. D.,
Fabbiano, G., Gilmore, D. \& Waller, W. H. 1995, \apj, 448, 98

\bibitem [Heckman 1997]{h97} Heckman, T. M. \& Leitherer, C. 1997, \aj, 114, 69

\bibitem [Jenkins 1987]{jenkins87} Jenkins, E. B. 1987, in Interstellar Processes, eds. D. J. Hollenbach \& H. A. Thronson, Jr., D. Reidel Publishing Company, 533

\bibitem [Heiles 1990]{h90} Heiles, C. 1990, \apj, 354,483

\bibitem [Kamphuis 1991]{k91} Kamphuis, J., Sancisi, R. \& van der Hulst, T.
1991, \aap, 244, 253

\bibitem [Lu et al. 1994]{lu94} Lu, L., Savage, B. D. \& Sembach, K. R. 1994, \apj, 437, L119

\bibitem [Lu et al. 1998]{lu98} Lu, L., Sargent, W.L.W. \& Barlow, T. A. 1998,
\aj, 115, 55


\bibitem [Mathewson and Ford 1984] {mat84} Mathewson, D. S. and Ford, V. L. 1984, Structure and Evolution of the Magellanic Clouds, eds. S. van den Bergh \& K. S. de Boer, Dordrecht:Kluwer, 25

\bibitem [Melnick et al. 1985]{mel85} Melnick, J., Moles, M. \& Terlevich, R. 1985, \aap, 149, L24

\bibitem[Meurer et al. 1992]{meu92} Meurer, G. R., Freeman, K. C., Dopita, M. A. and Cacciari, C. 1992, \aj, 103, 60

\bibitem[Meurer et al. 1995]{meu95} Meurer, G. R., Heckman, T. M., 
Leitherer, C., Kinney, A. Robert, C. \& Garnett, D. R. 1995, \aj, 110, 2665

\bibitem[Morton 1991]{mor91} Morton, D. C. 1991, \apjs, 77, 119

\bibitem[]{} Pagel, B.E.J., Simonson, E. A., Terlevich, R. J. 
\& Edmunds, M. G. 1992, \mnras, 255, 325

\bibitem[]{} Pettini, M., Lipman, K. \& Hunstead, R. W. 1995, \apj, 451, 100

\bibitem[]{} Sahu, M. S. \& Blades, J. C. 1997, \apj, 484, L125

\bibitem[]{} Sandage, A. 1978, \aj, 83, 904

\bibitem[]{} Renzini, A. \& Voli, M. 1981, \aap, 94, 175

\bibitem[]{} Sofia, U. J., Cardelli, J. A. \& Savage, B. D. 1994, 
\apj, 430, 650

\bibitem [Spitzer and Fitzpatrick 1993] {spi93} Spitzer, L., Jr. \& Fitzpatrick, E. 1993, \apj, 409, 299

\bibitem [Wakker and van Woerden 1991]{wakker91} Wakker, B. P. \& van Woerden, H. 1991, \aap, 250, 509

\bibitem [Wheeler et al. 1989]{wheeler89} Wheeler, J. C., Sneden, C. \& Truran, J. W. 1989, \araa, 27, 279

\bibitem [York et al. 1986]{york86} York, D. G., Dopita, M., Green, R. \&
Bechtold, J. 1986, \apj, 291, 627

\bibitem [York et al. 1990]{york90} York, D. G. Caulet, A., Rybski, P., Gallagher, J., Blades, J. C., Morton, D. C. \& Wamsteker, W. 1990, \apj, 351, 412

\end{thebibliography}
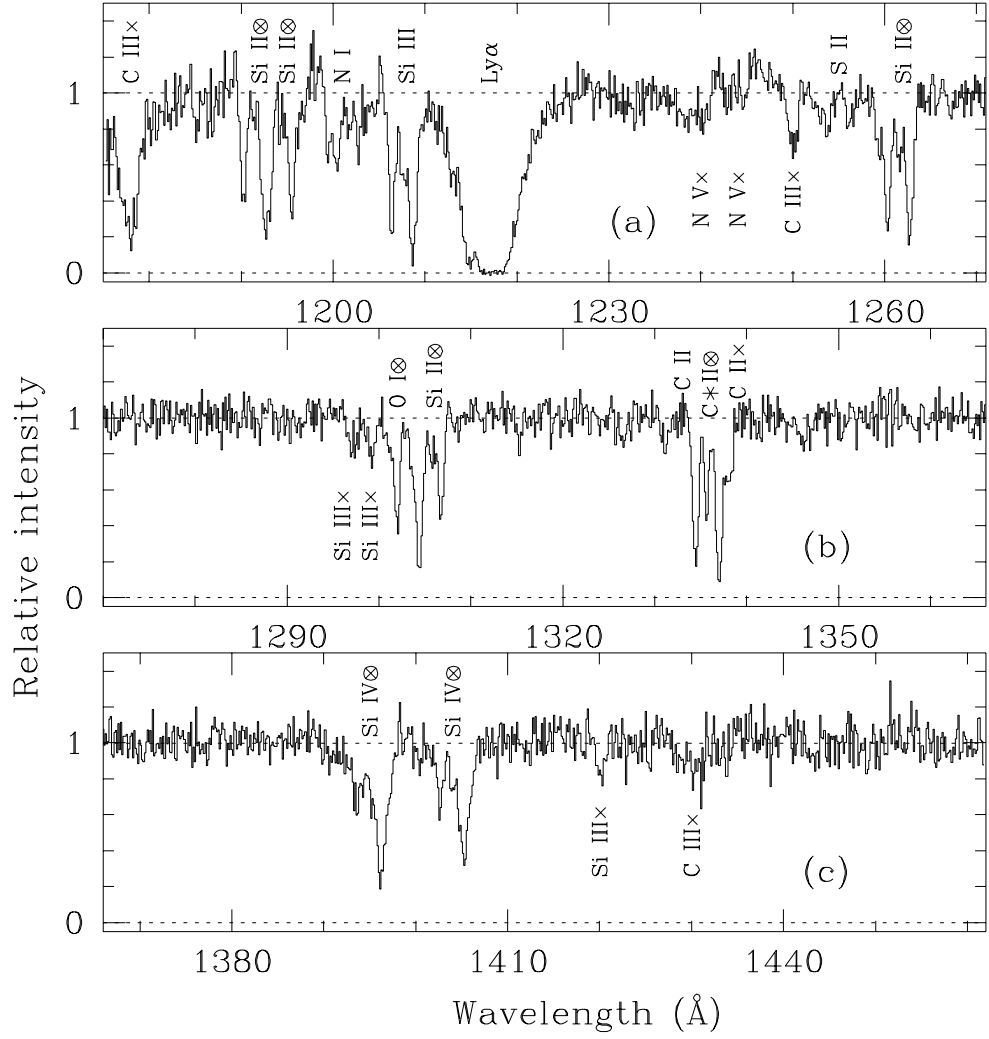
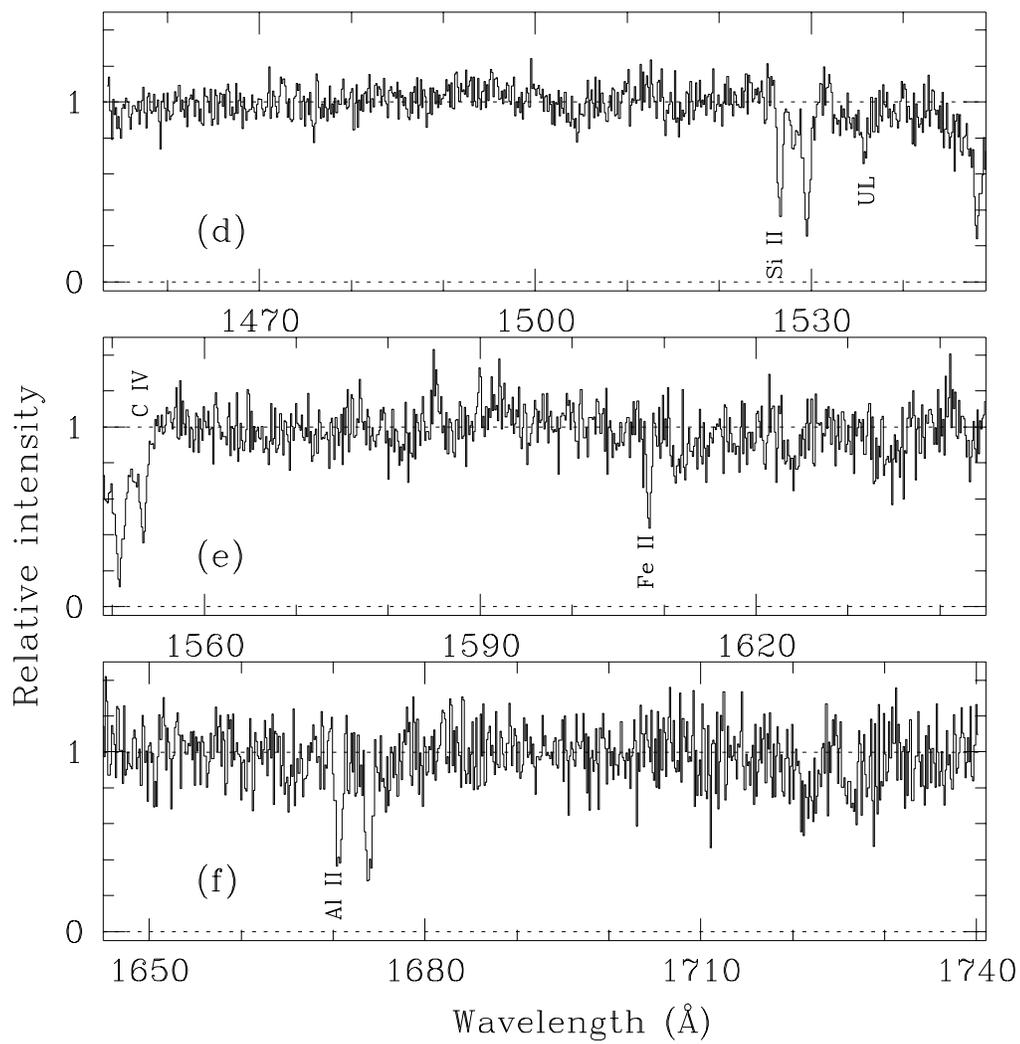
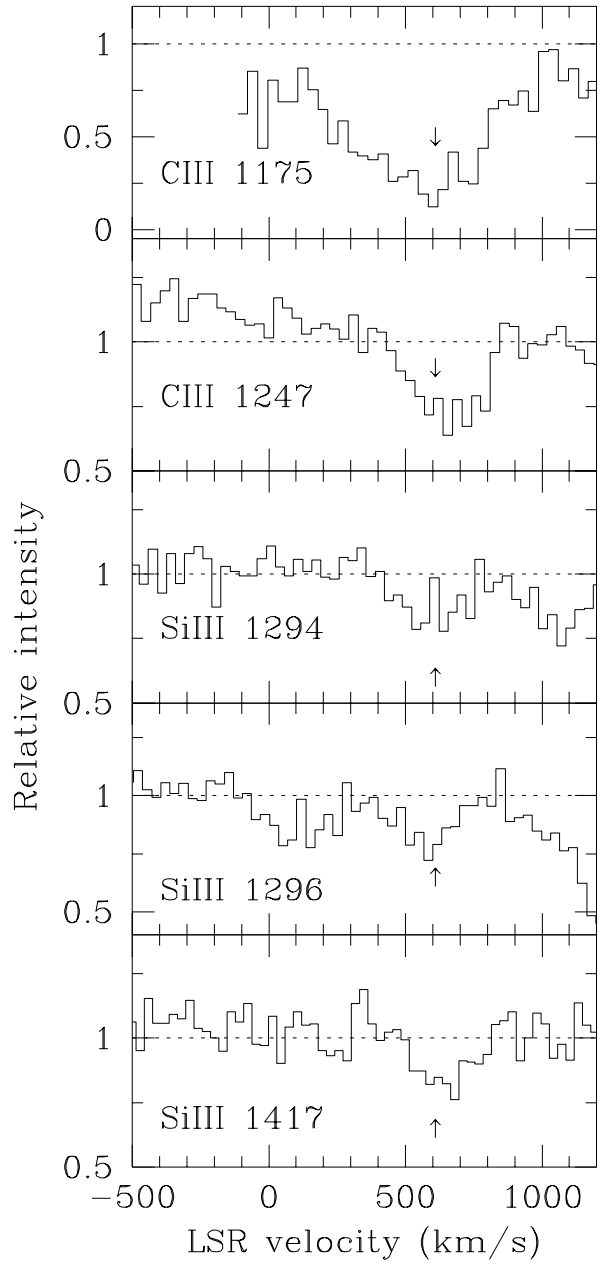
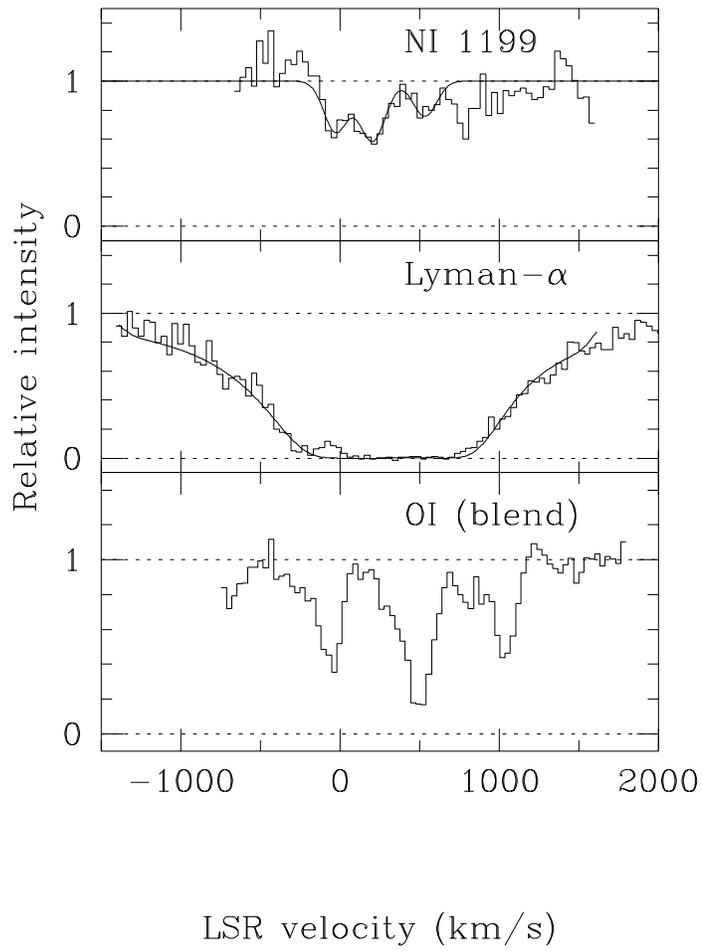
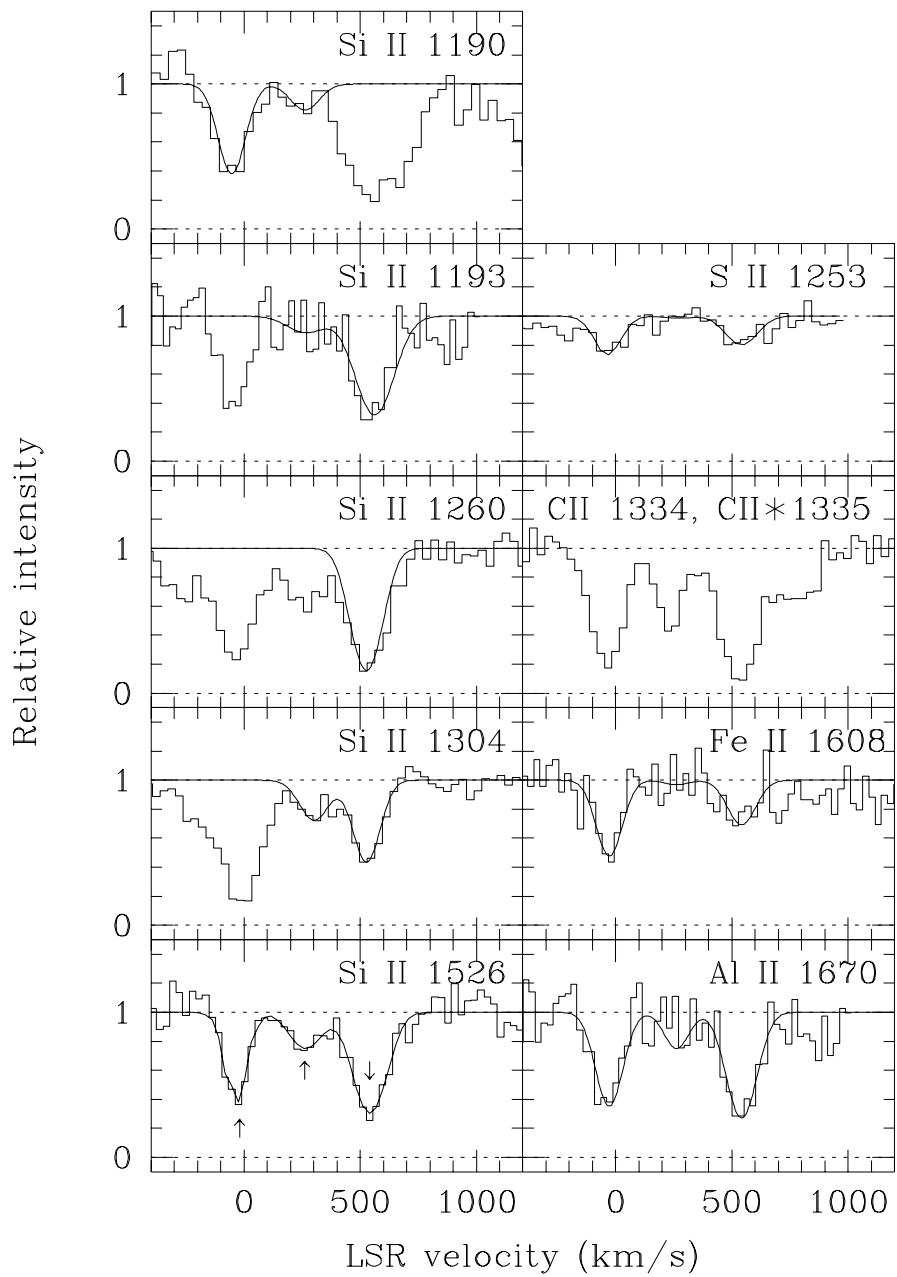
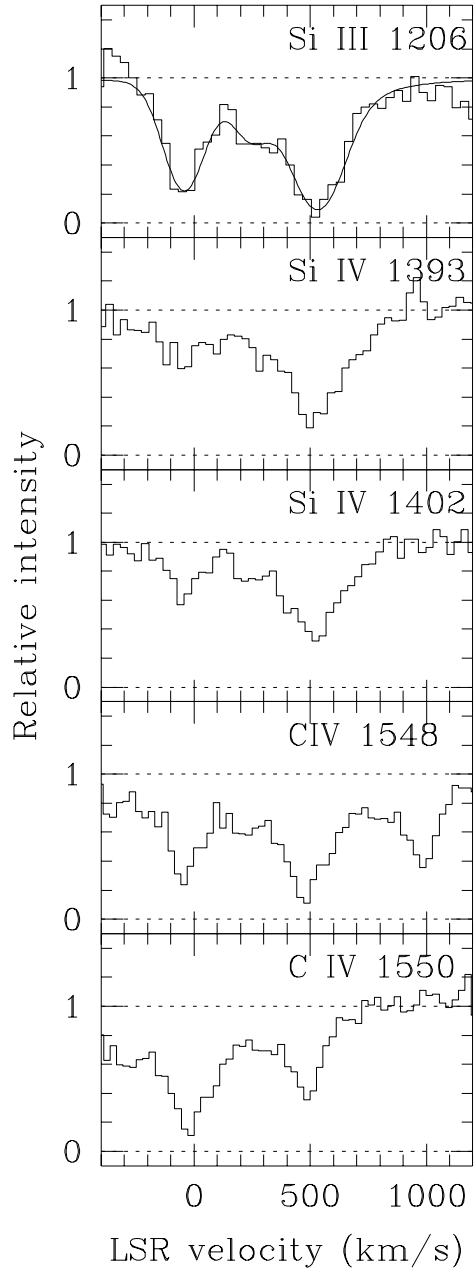
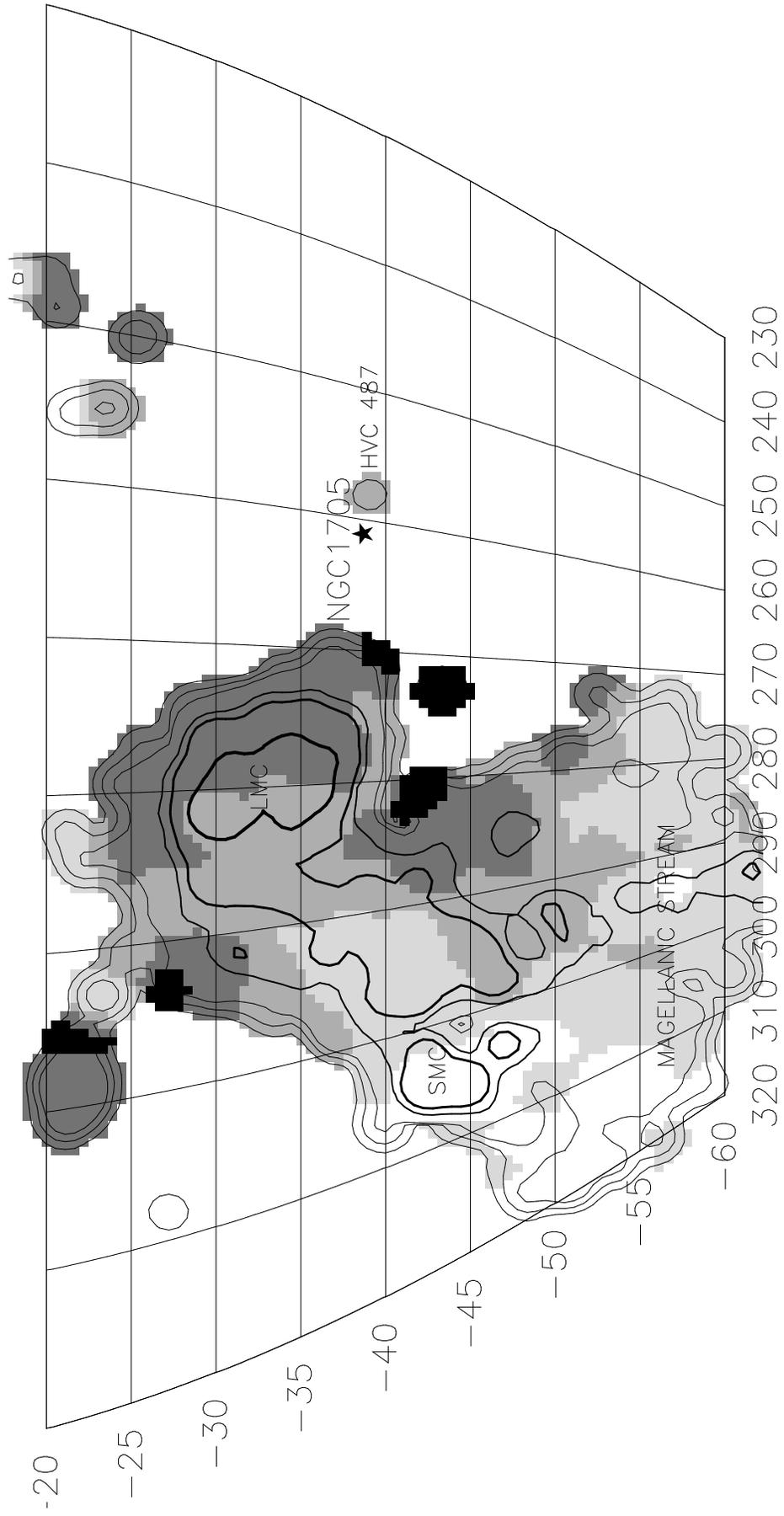
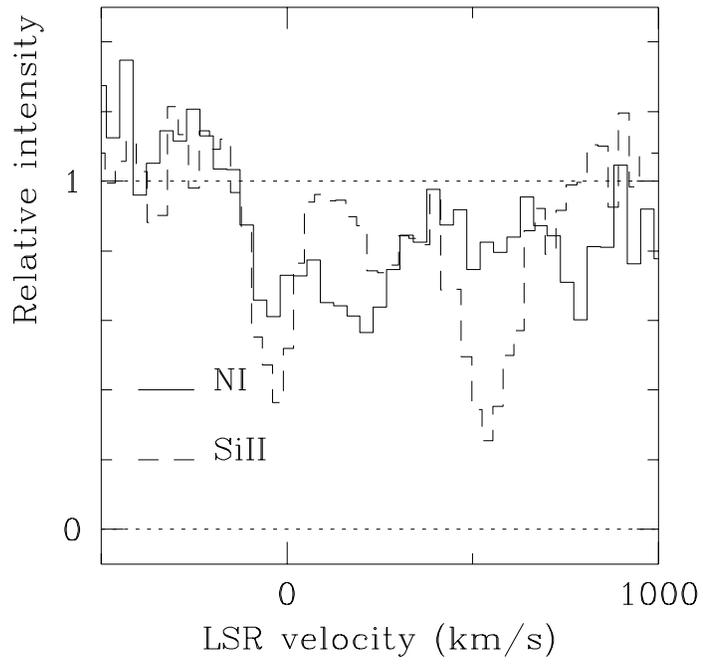
\end{document}